\documentclass[10pt, conference]{IEEEtran}

\usepackage[pdftex]{graphicx}
\DeclareGraphicsExtensions{.pdf,.jpeg,.png}

\usepackage[cmex10]{amsmath}

\usepackage[english]{babel}
\usepackage[T1]{fontenc}
\usepackage{inputenc}
\usepackage{subfigure}
\usepackage[numbers, sort]{natbib}
\usepackage{color}
\usepackage{fancyhdr}

\newcommand{\partAbl}[2]{\frac{\partial #1}{\partial #2}}

\begin{document}

\title{Interactive Data Exploration for High-Performance Fluid Flow Computations Through Porous Media}

\author{\IEEEauthorblockN{Nevena Perovi\'{c}, J\'{e}r\^{o}me Frisch, Ralf-Peter Mundani, and Ernst Rank}
\IEEEauthorblockA{Chair for Computation in Engineering\\
Technische Universit\"{a}t M\"{u}nchen, Germany\\
{Email: nevena.perovic@tum.de} }
}

%


\maketitle


\begin{abstract}

Huge data advent in high-performance computing (HPC) applications such as fluid flow simulations usually hinders the interactive processing and exploration of simulation results. Such an
interactive data exploration not only allows scientiest to `play' with their data but also to visualise huge (distributed) data sets in both an efficient and easy way. Therefore, we propose an HPC
data exploration service based on a sliding window concept, that enables researches to access remote data (available on a supercomputer or cluster) during simulation runtime without
exceeding any bandwidth limitations between the HPC back-end and the user front-end.

\end{abstract}


\pagestyle{empty}
\thispagestyle{fancy}
\lhead{}
\chead{}
\rhead{}
\lfoot{\copyright 2015 IEEE\\ 2014 16th International Symposium on Symbolic and Numeric Algorithms for Scientific Computing, Timisoara, 2014, pp. 463-470. doi: 10.1109/SYNASC.2014.68}
\cfoot{}
\rfoot{}

\flushbottom

%

\section{Introduction and Motivation}

In its 1987 report~\cite{McCormick1987}, the U.S.\ National Science Foundation stated \emph{computational steering} as valuable scientific discovery. Computational steering is the interaction
from a visual front-end with a running simulation (back-end) in order to modify properties such as geometry or boundary conditions for immediate feedback, i.\,e.\ allowing users to intervene with
an application during runtime and explore any effect caused by their changes \cite{Mulder1999}. According to~\cite{Marshall1990}, \emph{"steering enhances productivity by greatly reducing the
time between changes to model parameters and the viewing of the results"}. Even there has been a lot of research on steering within the last two decades, the combination of interaction and
high-performance computing (HPC) is still a challenging endeavour.

Typically, massive parallel simulation codes lead to huge data advent that cannot be interactively processed as necessary for a steering approach -- in our case the computed results of a fluid
flow simulation through porous media. Here, one major problem is the link between front- and back-end, i.\,e.\ the network with its bandwidth and latency restrictions. A proper paraphrase, to be
found in~\cite{Hoemmen2010}, states that \emph{arithmetic is cheap, latency is physics, and bandwidth is money}. Hence, to overcome this bottleneck special services are necessary in
order to leverage steering for demand-driven interactive HPC data exploration.

Therefore, we have designed a so-called \emph{sliding window concept}~\cite{Mundani2013} that allows users to `navigate' through data and retrieve computation results with respect to varying
regions of interest and resolutions. Key feature of that concept is to keep the total amount of data to be transferred between back- and front-end constant, hence any bandwidth limitations are
not exceeded at all times. Based on the sliding window, we have developed an HPC service for interactive data exploration, that was tested for parallel CFD applications. We will demonstrate
the applicability of this service for micro-scale simulations of flows through porous media and highlight the benefits for researchers in order to gain insight.

The remainder of this paper is as follows. First we will present the underlying physical fluid flow phenomena and their modelling in order to perform a high performance computation. We will then
introduce the proposed concept in order to integrate services into the previously described parallel fluid flow computations and present a complex engineering example of flows through porous
media. Finally, the paper will close with a short summary and outlook.

\section{High Performance Computing Concept for Computational Fluid Dynamics Simulations}
\label{sec:nse_and_hpc}


The computational fluid dynamics (CFD) simulations are based on an incompressible, isothermal, single-phase Newtonian fluid flow described by the conservation of mass equation written in
vector form as
\begin{equation}
\label{eq:cons_mass_incomp_div}
\nabla \cdot \vec u = 0 
\end{equation}
and three conservation of momentum equations
\begin{equation}
\label{eq:cons_mom_newtonian_fluid_incomp}
\partAbl{u_i}{t} + \nabla \cdot (u_i \vec u) = \nabla \cdot \left( \nu \nabla u_i \right) - \frac{1}{\rho} \nabla \cdot \left(p \vec e_i \right) + b_i 
\end{equation}
for $i \in \{x,y,z\}$, where $\vec u  = (u_x ~ u_y ~ u_z)^T$ denotes the velocity field in [m/s] in the three spatial dimensions $x,y,z$, $t$ the current time of the simulation in [s], $\nu$ the kinematic
viscosity in [m$^2$/s], $\rho$ the density in [kg/m$^3$], $p$ the pressure in [Pa], $\vec e_i$ the unity vector in the direction $i$, and $b_i$ sums up external volume forces such as acceleration
due to gravity in [m/s$^2$].

For the complete historical derivation of the full Navier-Stokes equations \eqref{eq:cons_mass_incomp_div} and \eqref{eq:cons_mom_newtonian_fluid_incomp} the interested reader is referred
to standard literature such as \cite{Batchelor2000, Ferziger2002, Hirsch2007}.

In order to solve the Navier-Stokes equations, a fractional step approach (also called projection method) proposed by Chorin \cite{Chorin1967} is applied. An intermediate velocity field is
computed while the pressure field is neglected at first. Using this intermediate velocity field, a pressure Poisson equation is established and has to be solved in order to correct and update the
intermediate velocity field and guarantee a convergence free velocity field at the next time step.

The numerical discretisation is based on a finite volume discretisation in space and a second order explicit Adams-Bashforth method in time \cite{Schwarz2011}. While applying Chorin's
projection method, the velocity terms are treated explicitly and the pressure terms are treated implicitly.

The backbone of the numerical discretisation and the basis for a parallel computation is formed by a highly adaptive data structure described in detail in \cite{Frisch2014SCPE}. As a matter of
completeness however, the principal points will introduced briefly.

\begin{figure}[!ht]
	\centering
		\includegraphics[width=0.31\textwidth]{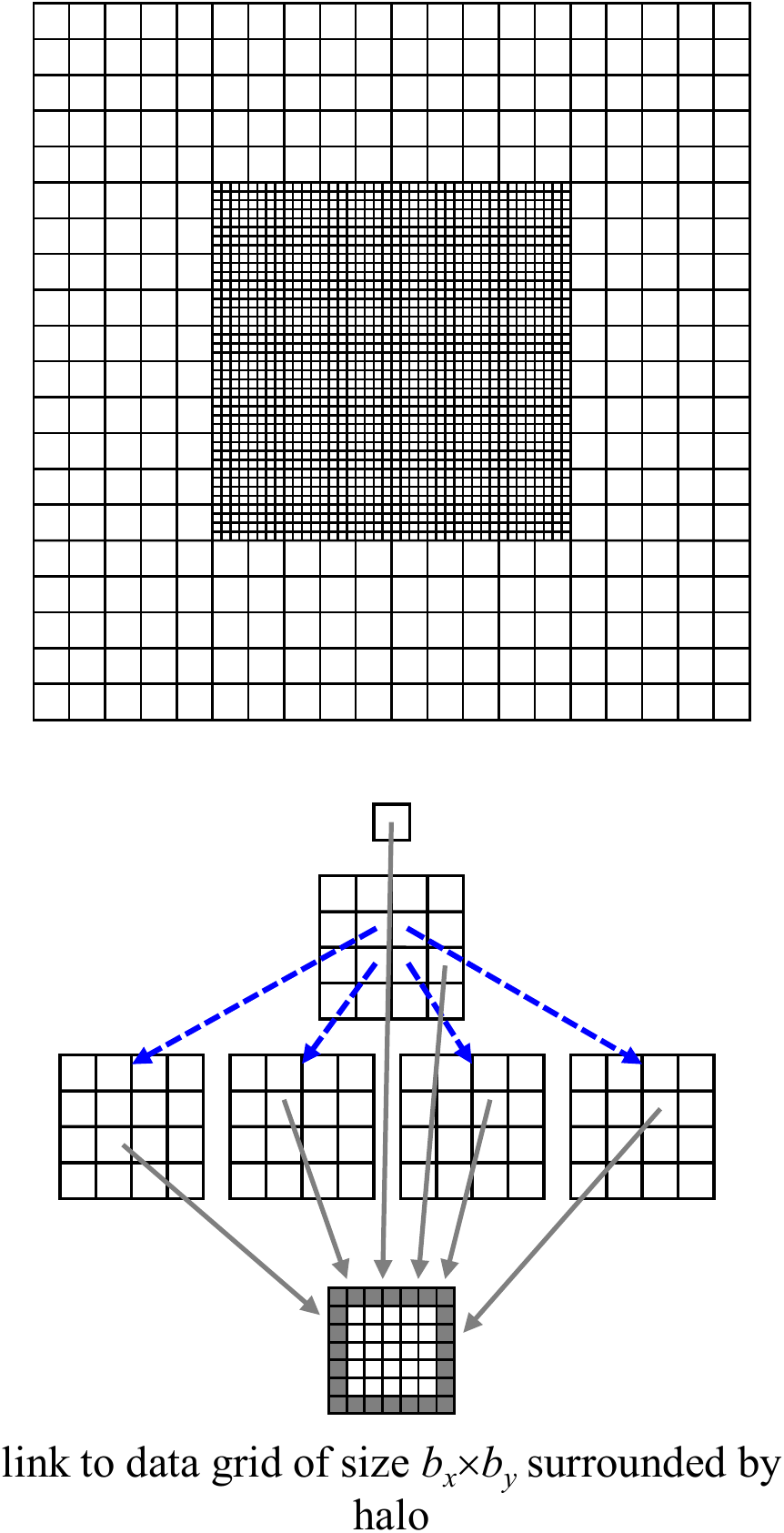}
	\caption{Top part: non-overlapping adaptive grid structure. Bottom part: hierarchical construction of the data structure out of logical grids and data grids.}
	\label{fig:data_structure}
\end{figure}

The data structure is constructed out of non-overlapping hierarchical block-structured orthogonal Cartesian grids and is depicted in figure \ref{fig:data_structure}. In the top part, the total
agglomerated computational grid can be seen, which will be distributed across different processes for performing a parallel computation in a distributed memory sense. In the bottom part, the
hierarchical construction is shown. In general, the data structure comprises two main parts: on the one hand, logical grids manage all hierarchical topological information such as links to parents
and children as well as geometrical information such as bounding boxes. On the other hand, data grids containing actual variables in order to perform numerical computations such as velocities,
pressures, temperatures etc.\ are linked to the previously mentioned logical management grids.

The hierarchical grid construction starts with a logical root grid at depth 0 which can be refined $r_x^t\times r_y^t \times r_z^t$ times. In case of the example shown in
figure~\ref{fig:data_structure}, four logical grids are refined again $r_x^s \times r_y^s \times r_z^s$ times which can be repeated recursively until the desired depth has been reached. The
refinement on the root level and on all subsequent lower levels can be selected arbitrarily in order to account for a non-cubic initial domain geometry.

Every logical grid is linked to one data grid containing all the variables which are necessary during the numerical simulation. A data grid is oblivious to topological and geometrical information and
relies on the link to the logical grid management structure. A data grid can also be called block and is an orthogonal Cartesian grid with the size of $s_x \times s_y \times s_z$. Every data grid is
identical and $s_i$ must be divisible without remainder by $r_i^s$ and by $r_i^t$ in order to avoid non-conforming setups.

The data grids are padded by one layer of ghost cells in all directions in order to perform a domain decomposition following a Schwarz approach \cite{Schwarz1870}. Hence, the data grids do
not overlap, but the ghost cells of the respective grids are overlapping. The solution process is repeated in an iterative manner containing two different phases: an update (or communication)
phase and a computation phase. During the update phase, boundary values are exchanged and updated in the corresponding neighbouring data grid's ghost cells. This process is repeated until
the solution has converged.

In case of a serial computation, the order between the update phase and the computation phase is given implicitly by the ordering of the lines of code in the program. In a parallel case using a
distributed memory approach no global time is present, and the synchronisation has to be implemented using explicit message exchanges. The order of the synchronisation calls is crucial, as a
wrong synchronisation procedure would result in undefined data states. The communication is handled by explicitly exchanging messages using the message passing interface (MPI). The
communication structure is divided into three phases (bottom-up, horizontal, and top-down) where a mixture of blocking and non-blocking communication calls ensures a correct predefined
ordering. Details about the communication structures can be found in \cite{Frisch2011Synasc}.

The implicit pressure Poisson equation is solved immediately on the grid itself rather than assembling one huge system matrix. As solution procedure a multi-grid-like solver integrated into the
data structure itself is applied. It uses the previously mentioned communication structures for its solving process and is able to treat large data amounts. Details about performance results are
published in \cite{Frisch2014SCPE}.

\begin{figure*}[t]
	\centering
		\includegraphics[width=0.80\textwidth]{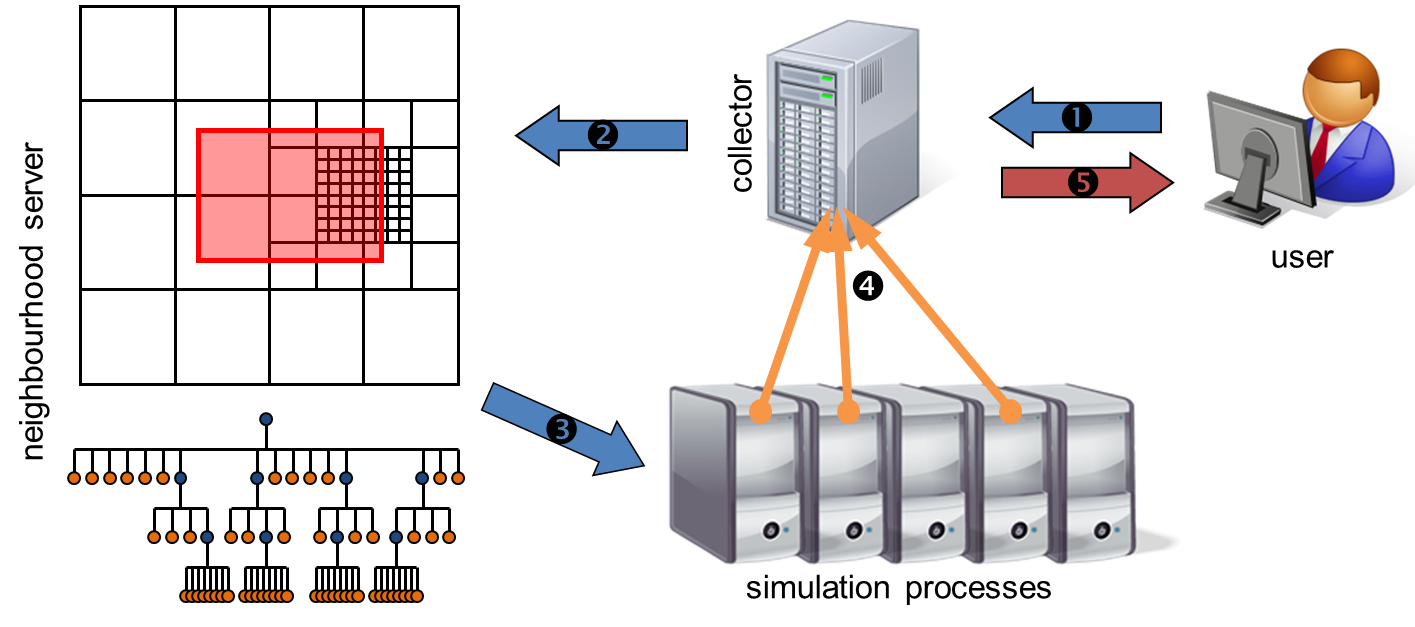}
	\caption{Service Based Concept -- User posts requests to collector (1) which are forwarded to the neighbourhood server (2). NBH identifies the simulation processes holding data grids and orders a visualisation (3). The data is then sent to the collector (4), which compresses them and forwards them to the user (5).}
	\label{fig:SW-NHC}
\end{figure*}

In addition to the computation processes containing the distributed grids, special dedicated processes are running concurrently containing a topological and geometrical repository of all data
grids over all processes without containing any actual data values such as pressures or velocities. These processes are called {\em neighbourhood servers} (NBH) as their primary task is to
answer queries such as \emph{`who is my neighbour on the east side?'}. Therefore, the NBH servers have a global view over the complete domain. A unique grid identification (UID) consisting
of a 64 bit integer ensures a specific addressing over different domains and different grids containing the encoded rank of the process as well as the grid ID on that specific process. If only one
neighbourhood server is used, every computing process has to communicate with the NBH server for querying information about neighbouring grids. This will lead at some point to a bottleneck.
The paper \cite{Frisch2012ISPDC} describes remedies using multiple NBH servers and special synchronisation techniques in order to overcome the bottleneck problem.

In order to get a good parallel performance, a load distribution has to be performed in order to distribute the amount of grids and, thus, the workload fairly over all available processes. To this
extent, space filling curves as introduced in \cite{Bader2013} can be used. The presented concept applies a Lebesgue curve \cite{Lebesgue1904} (also called z-curve due to its shape) as its
linearisation can be described and implemented easily by a Morton ordering \cite{Morton1966}.

Finally, {Frisch2014SCPE} shows performance measurements for the parallel computation of uniform as well as adaptive grid setups using all of the aforementioned concepts on the Blue
Gene/P installed at Universitatea de Vest din Timi\c{s}oara (UVT) in Romania.

\section{Service-Based Interaction Concept}
\label{sec:SB_Interaction_Concept}

Service-based interaction concepts get more and more important in nowadays computation environments. For the visualisation, the paper \cite{Mundani2013} introduces a so called {\em sliding window
concept} in order to overcome typical bandwidth problems while visualising large domains. Typically, CFD simulations generate giga bytes of data per time step which has to be treated and
visualised somehow. The typical state of the art procedure is to dump data for each time step to the supercomputer's network file system (or hard disk), transfer it completely to a front-end
visualisation machine, store it there, and explore and evaluate it in a visualisation environment in order to gain valuable scientific information, as `the purpose of computing is insight, not
numbers\footnote{written by Richard Wesley Hamming in the 1962 preface to {\em Numerical Methods for Scientists and Engineers.}}'. Unfortunately, this procedure puts a lot of strain on the
memory resources as well as the network bandwidth. A sample computation which was done with the above mentioned program using 80 billion cells on SuperMUC (one of Germany's national
supercomputers) had a memory footprint of over 28 tera byte per time step. Even if this information can be stored on disk, no visualisation program will be able to open and treat the contents
interactively.

Thus, a different approach was chosen. The user selects a window (i.\,e.\ region) of interest and together with a given network bandwidth this request is sent to the simulation program. The
program then selects the relevant parts by intersecting the window with the computational domain and checks how many information it can put inside the stream until the given bandwidth is
exceeded and sends the results back to the user who can visualise it on-the-fly. Using this approach, a user can either have a look at the complete domain in a coarse resolution or zoom into a
particular part and see it with all its details. The numerical computation is not influenced by the selection of the view itself; it is always performed on the finest possible resolution given by the
data structure and the previous problem definition.

Figure \ref{fig:SW-NHC} illustrates this request and selection service. A user is requesting to view a certain area with a certain resolution which his network connection can support without
problem and sends this request to a dedicated collector server (arrow 1). The collector contacts the neighbourhood server (arrow 2) and transmits the required information.

Once the NBH server has identified the processes which contain parts of the domain the user requested, it contacts them (arrow 3) and orders these processes to send the required information 
to the collector (arrow 4). The collector receives the information, compresses them in a serialised stream and sends them to the user (arrow 5). In case of a slow communication between
collector and user, the simulation processes are not blocked as they continue computing as soon as they have sent their information package to the collector.

Thus, the collector server represents an intermediate step separating the user from the computation processes in order not to disturb the simulation itself. Furthermore, it technically decouples
the computation processes' MPI communication from the user's socket based communication and acts as a collector and gateway to the outside world. Hence, the complete simulation back-end
can run on a supercomputer and the user connects at an arbitrary time to the collector. It is not necessary to be connected at the starting time of the simulation. Furthermore, a user can connect
or disconnect to or from the same simulation more than once. Thus, a user can check the progress of the simulation at multiple points in times while it is still running without the necessity of
interrupting the simulation or predefining the visualisation steps a-priori. This visualisation based service showed very promising results so far and will be used for porous media flows next.

Instead of ordering now visualisation tasks for interactive data explorations, different kinds of requests can be send to the collector and interpreted accordingly, such as changing of boundary
conditions, or refinement of grids. The complete simulation process can be driven by user requests submitted to a back-end simulation framework running on a supercomputer and acting as a
service handler.

\section{Application Example: Ground Water Flow through Porous Media}

The following section of the paper will describe a complex engineering application example treating a flow through a porous media using the above mentioned program concepts and services.

In order to describe a fluid flow through an arbitrary porous media, scientists rely on mathematically established and experimentally proven concepts established by Henry Darcy, a French
hydraulic engineer of the 19th century. According to his formulation, the ground water flow is directly proportional to the hydraulic gradient over a measured domain $i=\frac{\Delta p}{\Delta L}$
and a hydraulic permeability $k$ of earth layers through which the flow circulates. 

The hydraulic permeability $k$ carries within itself a vast amount of information about the physical properties of an investigated medium such as the consistency and distribution of soil particles,
the order of soil layers assuming an existence of non-homogeneous scenarios, and the type of soil (in a comparison to already experimentally determined values for several hundred types of
soil). Due to a high impact of $k$ on the accuracy of other hydraulic parameters in a wide range of engineering applications, a lot of effort is put into a precise definition, measurement, and
interpretation of the permeability.

The detailed description of experimental apparatuses which can be used to measure hydraulic conductivity $K$ and, thus, indirectly a permeability $k$ will not be presented here, as the relation
between $k$ [m$^{2}$] and $K$ [m/s] is already well established and consists only of known properties of a fluid, namely, viscosity, density, and acceleration due to gravitational forces.
Furthermore, the settings of physical experiments are highly dependable on the specific goal of examiners such as global values of permeability in one phase flow or particular values for every
single flow-phase and for every given soil layer. Due to the complexity of the measurement, we will not attempt to incorporate the complete procedure here, but the interested reader is referred to
already existing literature such as \cite{Method9100}.

In recent years, the ever increasing performance of computers provided scientists many possibilities to conduct very fine calculations on rather small domains. Hence,
current researches often address the coupling of the Navier-Stokes equations (micro scale) with the Darcy equations (macro scale) in order to bridge scales and to use micro scale results within
macro scale computations. Any call to the micro scale (as small part of the macro scale domain) requires a full 3D run of the simulation code, which is costly in terms of time and, thus, should
be executed only when necessary, i.\,e.\ demand-driven due to user requests.

Rather than making general conclusions about advantages and disadvantages that arise from the common concept of these two models, we have intentionally chosen the permeability variable
as an obvious link between them and based our research on how this value can be evaluated in one model and then without restriction used in another one. The complete study is conducted with
respect to the following assumptions: 
\begin{enumerate}
\item[a)] the granular pore geometry is highly irregular, without a specific pattern that repeats itself throughout the domain,
\item[b)] the irregularity of the domain is defined by a realistic granulometric sand curve,
\item[c)] a laminar flow regime is used for both models by fixing the Reynolds number to a low value,
\item[d)] the fluid is treated as incompressible,
\item[e)] the dominant flow direction is horizontal, i.\,e.\ parallel to the $x$-direction of the Cartesian coordinate system,
\item[f)] the influence of the gravity force as a body force is negligible compared to the viscous and pressure terms, but will be taken into consideration, in order to be consistent in the
formulation.
\end{enumerate}

\subsection{Governing Equations: Darcy Law}
\label{sec: equations}
In his experimental work Henry Darcy established a very important relationship among several hydraulic parameters which became the basis for vast amount of
subsequent theoretical studies. Hence, the water seepage through naturally settled ground layers $Q$ [m$^{3}$/s] is directly proportional to the hydrostatic pressure difference and the
permeability, and inversely proportional to the length of the sample and to the kinematic viscosity of the fluid. The Darcy equation has the following form:

\begin{equation}
\frac{Q}{A} = -\frac{k}{\mu} \cdot \frac{\Delta H_{p}}{L}\,,
\end{equation}

\begin{equation}
K = \frac{k \rho g}{\mu}\,,\quad\text{and}
\end{equation}

\begin{equation}
U_{darcy} = -K \frac{\Delta P}{L}\,,
\end{equation}

where the ratio $Q/A$ is often referred to as Darcy flux, which should not be confused with the fluid velocity $\vec u$ [m/s]. $k$ denotes the permeability of the chosen medium, $\mu$ the
dynamic viscosity of the fluid [kg/(m$\cdot$s)], $K$ the hydraulic conductivity [m/s], and $\Delta H_{p}$ [Pa] the pressure drop on the length $L$ [m] of the sand bed.

As the calculation and the visualisation of results are essentially conducted in all three directions, the corresponding vector components of the velocity are linked to the respective values of
the pressure drop, permeability values $\left[ k_{x}, k_{y}, k_{z} \right]$, and length differences $\Delta x, \Delta y$, and $\Delta z $.


\subsection{Governing Equations: Navier-Stokes Equations}

The Darcy Law is derived from physical Navier-Stokes equations introduced in section~\ref{sec:nse_and_hpc} which describe the fluid flow on a micro scale.

\subsection{Simulation Setup}

\begin{figure}[!bt]
\centering
\includegraphics[width=0.475\textwidth, angle=0]{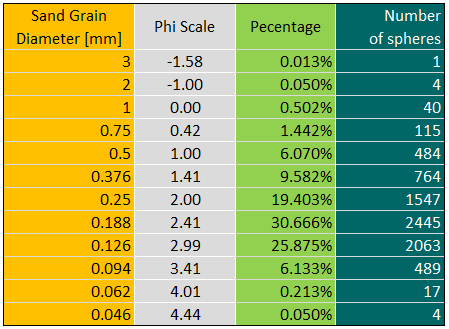}
\caption{Granulometric curve along with calculated number of spheres, extracted from the regular domain in order to form a fluid flow domain}
\label{fig:distribution}

\vspace{0.6cm}

\includegraphics[width=0.475\textwidth, angle=0]{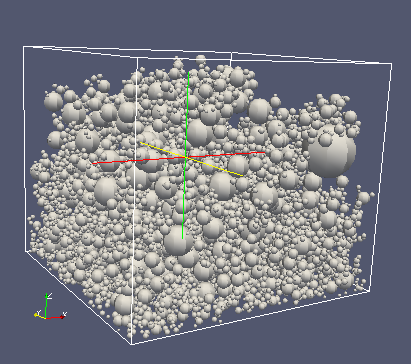}
\caption{Complex geometry generated using random distribution function (stage 1). Inversion of this geometry forms a fluid flow domain (stage 2) }
\label{fig:cc2}
\end{figure}

In order to be able to evaluate the obtained results, the same pore geometry is used in a set of different scenarios, in which the value of porosity is defined as constant for a single scenario.
Such restrictions led to unambiguous calculation settings, defined in figure~\ref{fig:distribution}, which are then suitable for subsequent analyses.

Furthermore, in this chapter two cases based on different geometries will be presented, with calculated values of permeability, challenges that we meet, description of ongoing studies as well as
how the previously introduced service-based exploration concept can be used throughout the process, in order to analyse our data sets.

\subsection{Generation of the Domain}

The generated testing domains can be split into two groups: 
\begin{itemize}
\item[1.] with random, irregular geometry, that can be seen in figures~\ref{fig:cc2} and \ref{fig:irreg_geom} and
\item[2.] with regular geometry setup, which is thoughtfully tested, but not presented here, as the first irregular case offers a far better basis for exploration of the introduced sliding window concept. 
\end{itemize}
Whereas all regular domains are generated analytically, taking into consideration only geometrical parameters of inserted spheres, irregular domains are far more complicated and generated
as a rectangular shape with predefined dimensions, in which all spheres with different diameters representing soil grains according to the granulometric distribution (cf.\
figure~\ref{fig:distribution}) are placed using random distribution functions. Geometric intersection tests are performed in order to prevent the overlapping of spheres, such that the generated
sample has a physical validity.

Once the sample geometry is generated and described using a boundary representation method, a block-mesh generator based on an octree generation as described in \cite{Mundani2006} is
creating a 3D volume discretisation of the domain by setting the corresponding boundary conditions on the data grids.

Due to agglomerations of `artificial' sand grains near the walls, the results for the relevant parameters such as velocity, pressure, and permeability values were very disperse. In order to reduce
local boundary influences, subsequent analyses are conducted on sub-domain samples closer to the centre of the main domain.

The amount of solid particles goes from only 35\% up to densily packed 65\% of the total volume. As the inflow velocity can have a significant influence on the velocity distribution throughout the
domain -- leading to possible numerical instabilities -- a larger simulation domain was intentionally set in $x$-direction, where a fixed boundary condition value was applied.

The implemented boundary conditions are defined as follows:
\begin{itemize}
\item at the left-most boundary -- Dirichlet boundary conditions for the inflow velocity with a fixed value of $u_{in,x} = 1~m/s$, $u_{in,y} = u_{in,z} = 0$ and homogeneous Neumann boundary
conditions for the pressure,
\item at the right-most boundary -- free outflow boundary conditions with homogeneous Neumann boundary conditions for the velocity and Dirichlet conditions with a pressure value set to zero,
\item at all other walls of the domain -- No slip boundary conditions with Dirichlet boundary conditions fixed at zero for the velocities (i.\,e.\ the velocity of the fluid is at rest in the vicinity of the wall)
and Neumann boundary conditions for the pressure, and
\item at the surface of sand grains -- No slip boundary condition, as all spheres are treated as solid (similar to the walls), with fixed position over the space and non-deformable over the time
particles.  
\end{itemize}

\begin{figure}[ht]
\centering
\includegraphics[width=0.48\textwidth, angle=0]{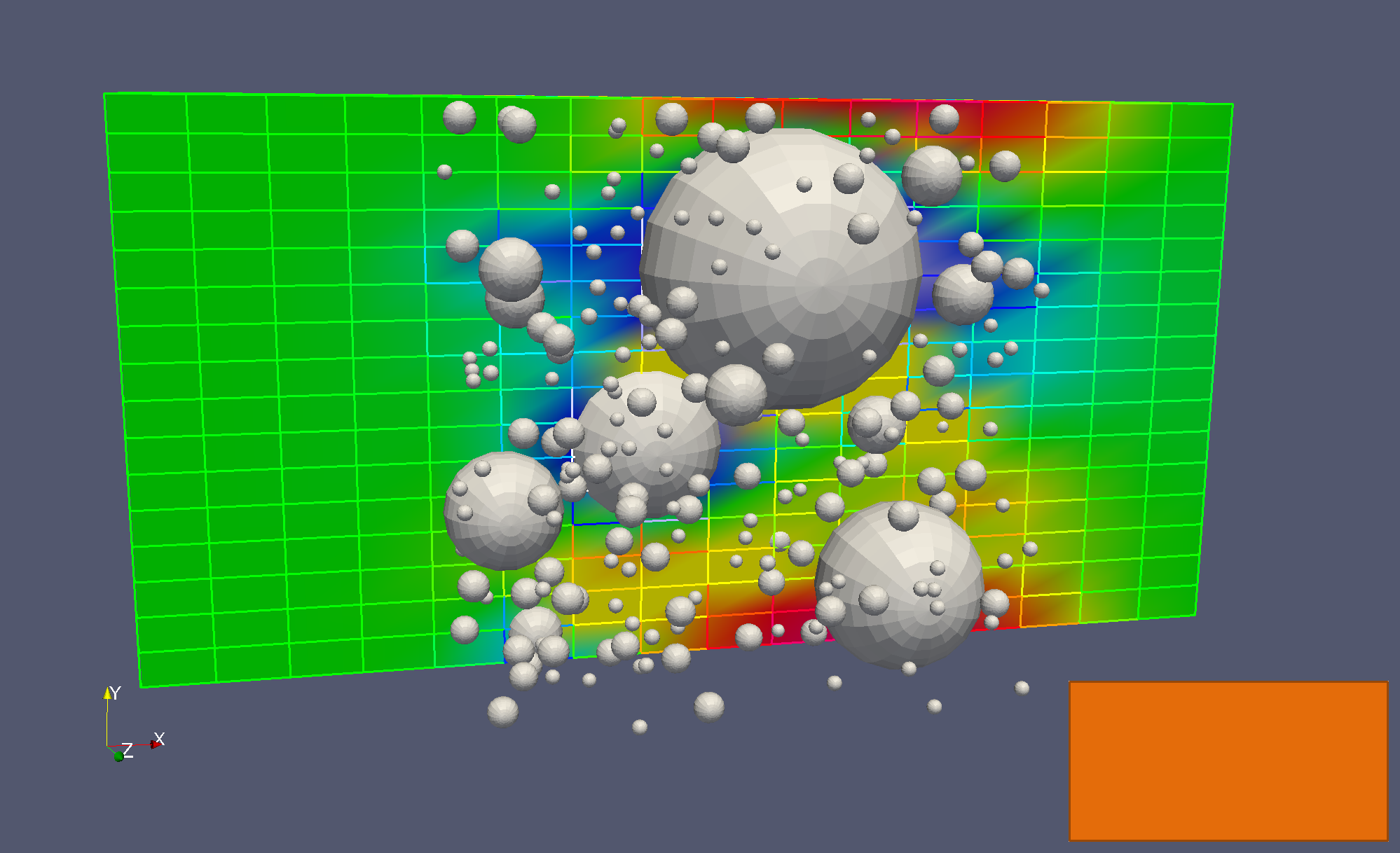}\\
\vspace{0.25cm}
\includegraphics[width=0.48\textwidth, angle=0]{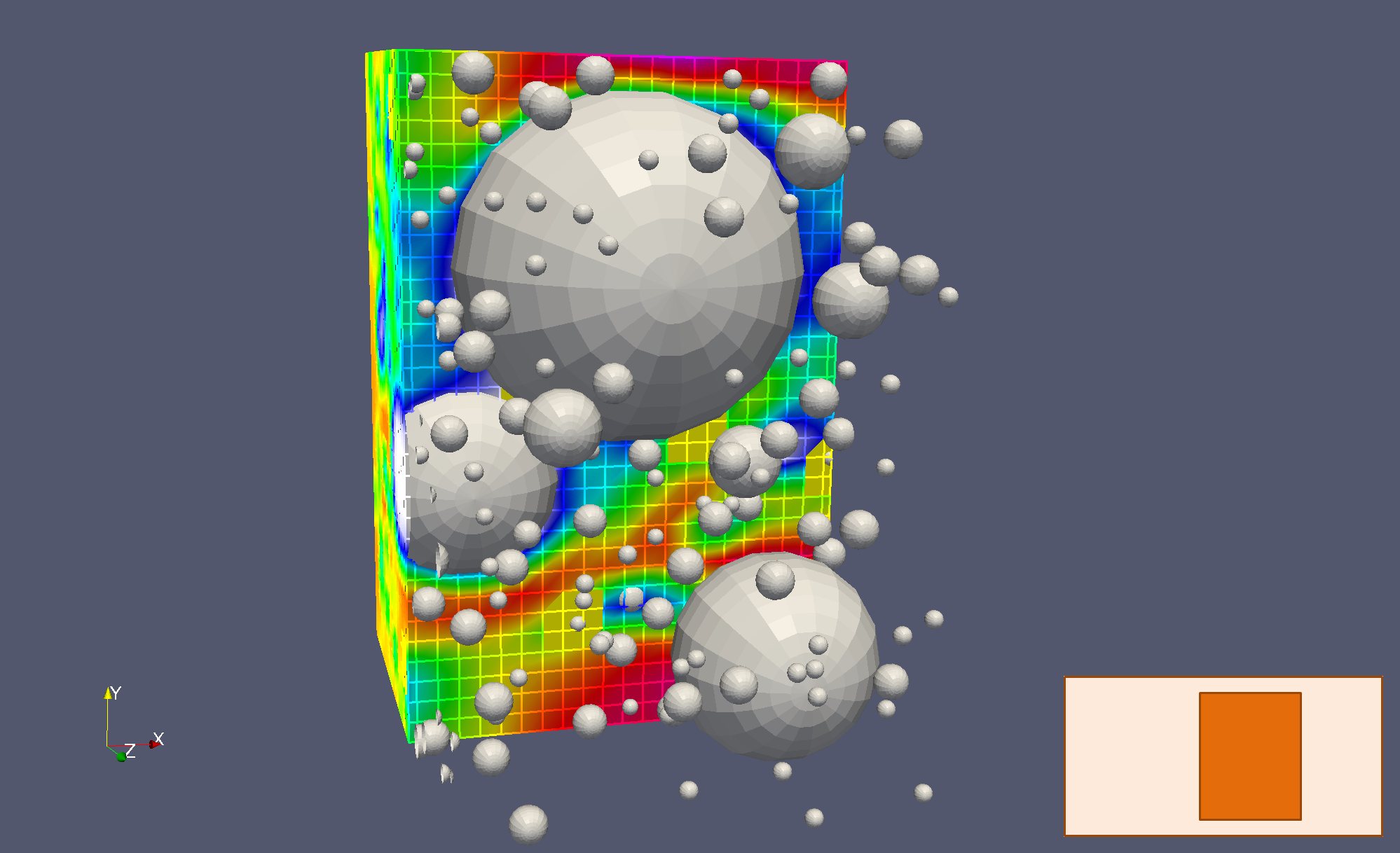}\\
\vspace{0.25cm}
\includegraphics[width=0.48\textwidth, angle=0]{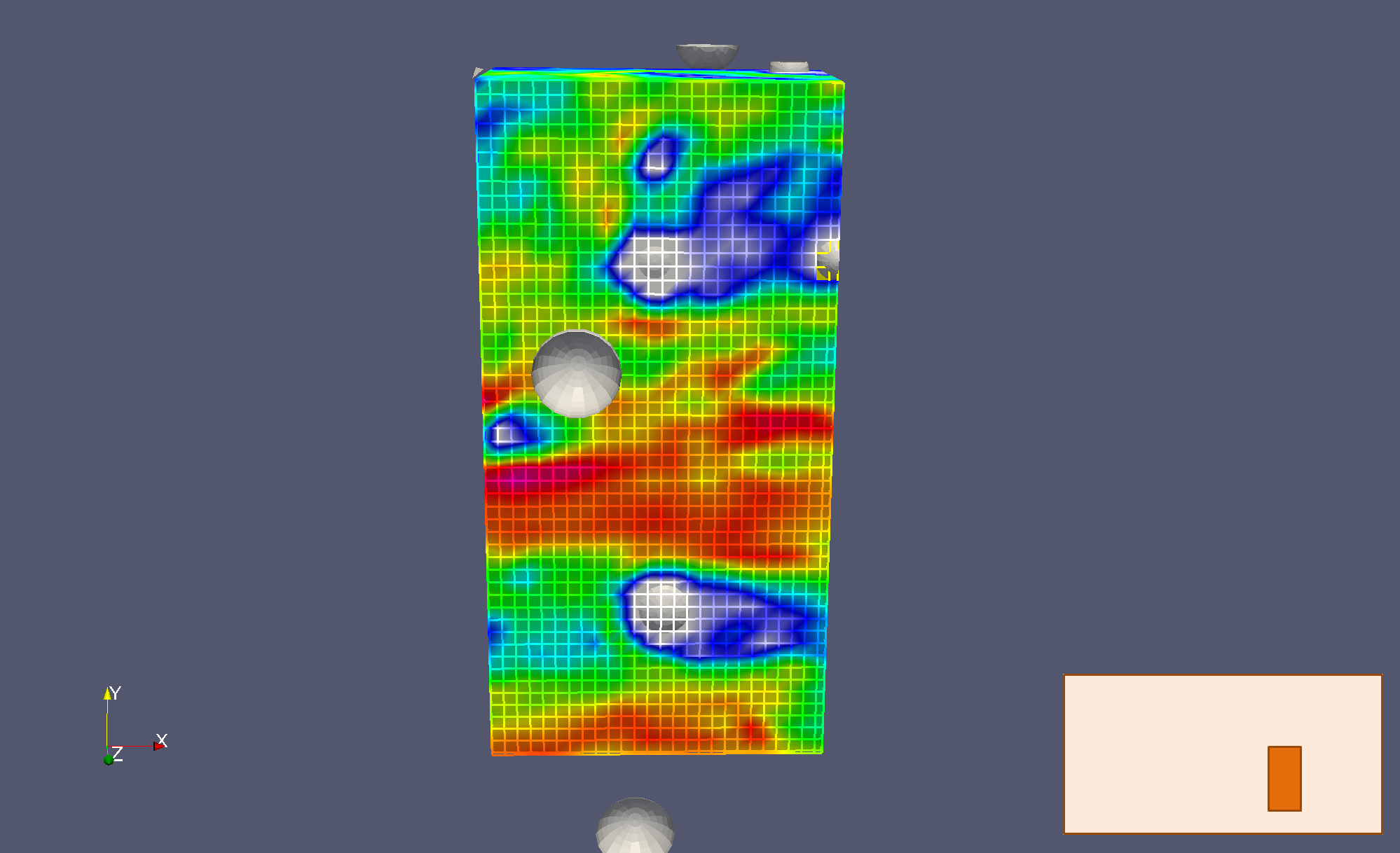}
\caption{Interactive multi-level domain exploration using the sliding window concept. Displayed are velocity magnitudes on a selected region of interest. In the top picture a `global' view is selected giving a global impression. In the middle and the lower picture, more and more details can be seen while zooming in.}
\label{fig:sliding_window_example}
\end{figure}

Figure \ref{fig:sliding_window_example} highlights the benefits of on interactive data exploration using the sliding window concept for a complex flow through a porous media. The simulation is always executed on the deepest level of refinement (in this case three levels) using around 6 million cells. On the top level (top part of figure  \ref{fig:sliding_window_example}) only a few cells are visualised, but they can give a good impression of the overall flow behaviour using a very limited amount of the available, i.\,e.\ computed information.

Moreover, the smallest details would not be distinguishable any more as the differences in scales are too big. If more details should be visualised, only one small part of the geometry is visualised in more details (middle part of figure  \ref{fig:sliding_window_example}). Here, more details can be analysed around a zoomed area visualising more or less the same amount of cells and, thus, keeping the transmitted bandwidth the same as for the top level visualisation. The initial geometry was overlaid in all of the examples in order to get an idea of the zoomed region. Even more details can be seen on a smaller region in the lower part of figure  \ref{fig:sliding_window_example}.

\subsection{Further analyses}
\begin{figure}[ht]
\centering
\includegraphics[width=0.475\textwidth, angle=0]{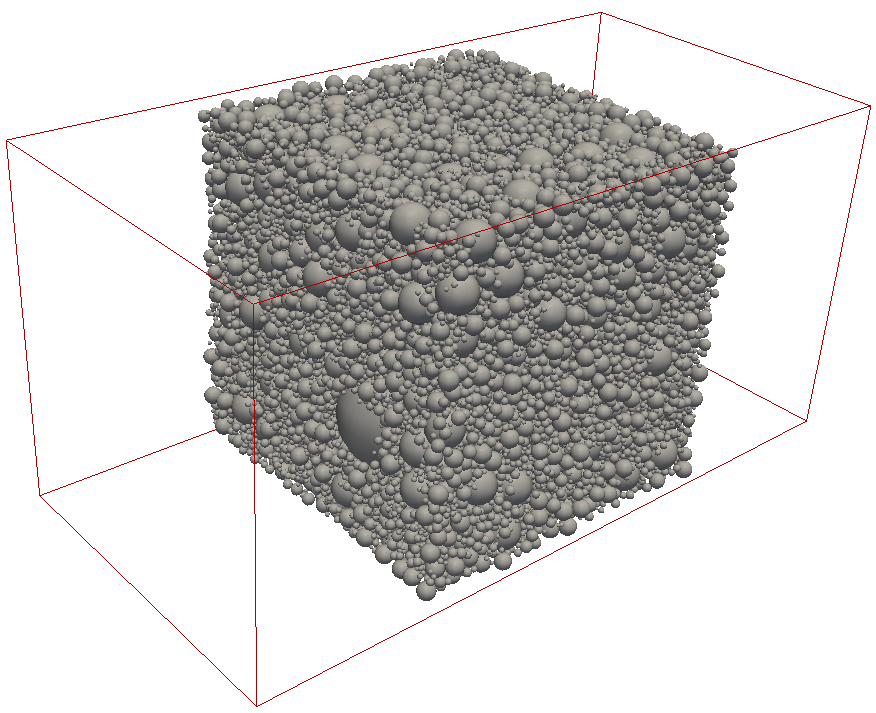}
\caption{Calculation domain randomly filled up with spheres of different size, with predefined volumetric contribution of all fractions to the total volume of solid phase}
\label{fig:irreg_geom}
\vspace{0.5cm}
\includegraphics[width=0.475\textwidth, angle=0]{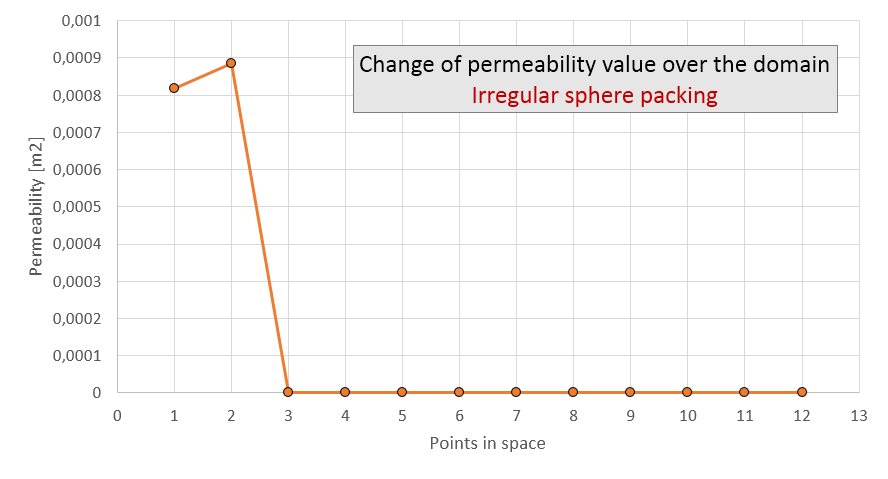}
\caption{Calculated permeability value for 12 arbitrarily chosen points throughout irregularly (randomly) set domain. Two first points are taken from the part of the domain, where no spheres are
positioned}
\label{fig:irreg_1}
\vspace{0.75cm}
\includegraphics[width=0.475\textwidth, angle=0]{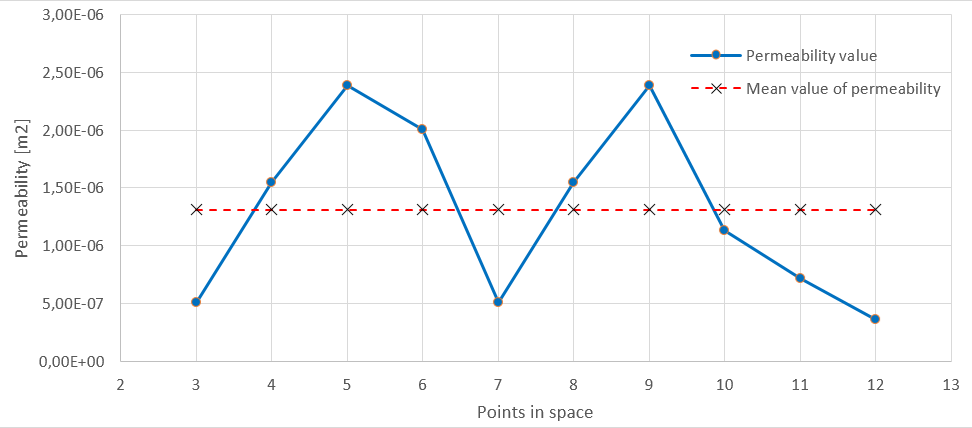}
\caption{Calculated and mean value of permeability for 10 points solely from part of the domain filled with spheres -- irregular sphere packing}
\label{fig:irreg_2}
\end{figure}

\vspace{0.2cm}
Taking into account the previously described concept and all benefits that arise from its implementation, different parameters' analyses can be carried out already after first steps of the simulation process, giving us an overall trend of the physical validity of the conducted experiments. One of these examples is depicted in the figures \ref{fig:irreg_geom}, \ref{fig:irreg_1}, and \ref{fig:irreg_2}, where the fluid flow simulation through the generated porous medium is presented with depicted sensitivity analyses of permeability value, previously denoted in subsection \ref{sec: equations}.  \\
Having on mind the complexity of the generated geometry and the complexity of the data structure that supports such cases, analysing the complete data set in real time would be demanding task, if at all possible to transmit such huge amounts of data over the underlying network, without having side effects. Furthermore, for the most such analyses, the whole data set is not really necessary for studying a specific parameter in a specific part of the domain under particular circumstances. This concept allow us to conduct memory and time consumption saving tests without any loss of data that could influence the results itself.

In figure \ref{fig:irreg_geom} randomly generated highly irregular geometry, which was investigated in this particular study is shown. On the deepest level of our data structure (resulting to the resolution of $320 \times 160 \times 160 $ cells) more than 8 million cells are taken into calculation. Analysing parameters at this level could be cumbersome in the means of transferring calculated data for a particular desired time step. For that reason, the visualisation of the whole domain is done on the coarsest level, in order to check all set boundary conditions and possible conflicts that could be already there effectively spotted. On the next finer level, also depicted in figure \ref{fig:sliding_window_example} (middle plot), the general trend of permeabilty value is visualised and consequently shown in figure \ref{fig:irreg_1}, whereas on the deepest, in this case third level of refinement the detailed analyses are conducted, where necessary plots of spatial distribution of explored values are produced. One similar plot is shown in figure \ref{fig:irreg_2}.\\
Closely observing enclosed diagrams, it is obvious that some irregularities in behavior of examined parameter are not to be detected on the coarser level, show in figure \ref{fig:irreg_1}. If there were no possibility to explore mentioned parameter on a finer level, the shown behavior would lead to the wrong conclusion that our permeability value is constant throughout domain and representative for every spatial distribution of inserted spheres. Nevertheless, enclosed finer representation, depicted in figure \ref{fig:irreg_2} highlights the spatial fluctuation of permeability (drawn with blue line) comparing to calculated mean value (drawn with red dot line). Such aberration can be explained with fact that the domain was highly irregular with no patterns to be repeated throughout it and therefore the values of permeability are expected to be highly diverse. Ongoing research goes into direction of defining such relationship between aforementioned value and sieve curve (cf. figure \ref{fig:distribution}) on which basis the complex geometry was generated. Such defined permeability value should be completely independent of spatial distribution of solid bodies over the domain. Already gathered results are promising.   

Shortly explained setting of conducted analyses is just an example of application of our concept in diverse engineering problems. Nevertheless, it shows clearly, that such analyses are highly feasible for wide range of physical phenomena and their fundamental parameters, providing a powerful tool for  experts to either establish further new relationships among variables of interests or explore new behavior by interactively changing the set of input values on-the-fly, having immediate response and logically comprehensible feedback.

Along with the detailed explanation of the implemented high performance computing routines, some first obtained results are presented in this section, with intention to validate (in future) our finite
volume concept for porous media flow in comparative analyses to published results based on finite elements solvers as well as to data obtained from -- specifically for this purpose -- conducted
experiments.

\section{Conclusion}

In order to tackle the huge data advent arising from massive parallel computations, a sliding window concept based on services was introduced in this paper, enabling a user to select the
computational domain as a whole or highlight a specific detail and visualise it interactively.

This paper presented the basic concepts and data structures of the underlying computational domain as well as the necessary data exchange and load balancing strategies in order to execute
a simulation on a massive parallel computing systems. Hence, this paper combined the concepts of the sliding window visualisation with HPC computing in order to simulate and interactively
explore a complex engineering scenario such as a flow through porous media. The HPC service based approach shall be investigated further in future in order to obtain benefits from
on-demand computing and adaptive mesh refinement for instance.

\section{Acknowledgement} 

This publication is partially based on work supported by Award No. UK-c0020, made by King Abdullah University of Science and Technology (KAUST). Furthermore, the authors would like to
cordially thank for the support and usage of the Blue Gene/P at Universitatea de Vest din Timi\c{s}oara (UVT) in Romania. 


\bibliographystyle{IEEEbib}
\bibliography{paper}

\begin{thebibliography}{10}

\bibitem{McCormick1987}
B.H. McCormick, T.A. DeFanti, and M.D. Brown,
\newblock ``Visualization in scientific computing,''
\newblock {\em Computer Graphics}, vol. 21, no. 6, 1987.

\bibitem{Mulder1999}
J.D. Mulder, J.J. van Wijk, and R.~van Liere,
\newblock ``A survey of computational steering environments,''
\newblock {\em Future Generation Computer Systems}, vol. 15, no. 1, pp.
  119--129, 1999.

\bibitem{Marshall1990}
R.~Marshall, J.~Kempf, S.~Dyer, and C.C. Yen,
\newblock ``Visualization methods and simulation steering for a {3D} turbulence
  model of {L}ake {E}rie,''
\newblock {\em ACM SIGGRAPH Computer Graphics}, vol. 24, no. 2, pp. 89--97,
  1990.

\bibitem{Hoemmen2010}
M.~Hoemmen,
\newblock {\em Communication-avoiding {K}rylov Subspace Methods},
\newblock Ph.D. thesis, University of California at Berkeley, 2010.

\bibitem{Mundani2013}
R.-P. Mundani, J.~Frisch, and E.~Rank,
\newblock ``Towards interactive {HPC}: {S}liding window data transfer,''
\newblock in {\em Proc. of the 3rd Int. Conference on Parallel, Distributed,
  Grid and Cloud Computing for Engineering}. 2013, Civil-Comp Press.

\bibitem{Batchelor2000}
G.K. Batchelor,
\newblock {\em An Introduction to Fluid Dynamics},
\newblock Cambridge University Press, 2000.

\bibitem{Ferziger2002}
J.H. Ferziger and M.~Peri{\'c},
\newblock {\em Computational Methods for Fluid Dynamics},
\newblock Springer, 3rd rev. edition, 2002.

\bibitem{Hirsch2007}
C.~Hirsch,
\newblock {\em Numerical Computation of Internal and External Flows, Volume 1},
\newblock Butterworth--Heinemann, 2nd edition, 2007.

\bibitem{Chorin1967}
A.J. Chorin,
\newblock ``Numerical solution of the {N}avier-{S}tokes equations,''
\newblock {\em Mathematics of Computation}, vol. 22, no. 104, pp. 745--762,
  1968.

\bibitem{Schwarz2011}
H.~R. Schwarz and N.~K{\"o}ckler,
\newblock {\em Numerische {M}athematik},
\newblock Vieweg + Teubner, 8th rev. edition, 2011.

\bibitem{Frisch2014SCPE}
J.~Frisch, R.-P. Mundani, and E.~Rank,
\newblock ``Adaptive multi-grid methods for parallel {CFD} applications,''
\newblock {\em Scalable Computing: Practice and Experience}, vol. 15, no. 1,
  pp. 33--48, 2014.

\bibitem{Schwarz1870}
H.~A. Schwarz,
\newblock ``Ueber einen {G}renz{\"u}bergang durch alternirendes {V}erfahren,''
\newblock {\em Vierteljahrsschrift der Naturforschenden Gesellschaft in
  Z{\"u}rich}, vol. 15, pp. 272--286, 1870.

\bibitem{Frisch2011Synasc}
J.~Frisch, R.-P. Mundani, and E.~Rank,
\newblock ``Communication schemes of a parallel fluid solver for multi-scale
  environmental simulations,''
\newblock in {\em Proc. of the 13th Int. Symposium on Symbolic and Numeric
  Algorithms for Scientific Computing (SYNASC)}. 2011, pp. 391--397, IEEE
  Computer Society.

\bibitem{Frisch2012ISPDC}
J.~Frisch, R.-P. Mundani, and E.~Rank,
\newblock ``Resolving neighbourhood relations in a parallel fluid dynamic
  solver,''
\newblock in {\em Proc. of the 11th Int. Symposium on Parallel and Distributed
  Computing}. 2012, pp. 267--273, IEEE Computer Society.

\bibitem{Bader2013}
M.~Bader,
\newblock {\em Space-Filling Curves -- {A}n Introduction with Applications in
  Scientific Computing},
\newblock Springer-Verlag, 2013.

\bibitem{Lebesgue1904}
H.L. Lebesgue,
\newblock {\em Le\c{c}ons sur l'int{\'e}gration et la recherche des fonctions
  primitives},
\newblock Gauthier-Villars, 1904.

\bibitem{Morton1966}
G.M. Morton,
\newblock ``A computer oriented geodetic data base and a new technique in file
  sequencing,''
\newblock Tech. {R}ep., IBM Ltd., Ottawa, Ontario, Canada, 1966.

\bibitem{Method9100}
``Method 9100 -- {S}aturated hydraulic conductivity, saturated leachate
  conductivity, and intrinsic permeability,''
\newblock Tech. {R}ep., United States Environmental Protection Agency EPA,
  1986.

\bibitem{Mundani2006}
R.-P. Mundani,
\newblock {\em Hierarchische {G}eometriemodelle zur {E}inbettung verteilter
  {S}imulationsaufgaben},
\newblock Ph.D. thesis, Universit{\"a}t Stuttgart, 2006.

\end{thebibliography}

\end{document}